# Multilayer gas cells for sub-Doppler spectroscopy

## Azad Ch. Izmailov

*Institute of Physics, Azerbaijan National Academy of Sciences, H. Javid av. 33, Baku, Az-1143, Azerbaijan*

e-mail: azizm57@rambler.ru

**Abstract**

We have carried out theoretical research on ultra-high resolution spectroscopy of atoms (or molecules) in the suggested cell with a series of plane-parallel thin gas layers between spatially separated gas regions of this cell for optical pumping and probing. It is shown the effective velocity selection of optically pumped atoms because of their specific transit time and collisional relaxation in such a cell, which lead to narrow sub-Doppler resonances in absorption of the probe monochromatic light beam. Resolution of this spectroscopic method is analyzed in cases of stationary and definite nonstationary optical pumping of atoms by the broadband radiation versus geometrical parameters of given cells and pumping intensity. The suggested multilayer gas cell is the compact analog of many parallel atomic (molecular) beams and may be used also as the basis of new compact optical frequency standards of high accuracy.

## 1. Introduction

It is very important to elaborate effective methods of high-resolution spectroscopy which allow to analyze a structure of spectral lines hidden by the Doppler broadening because of a thermal motion of atoms or molecules [1]. Given methods are used also for creation of more perfect optical frequency standards [2].

In 1992-1993 new methods of sub-Doppler laser spectroscopy were theoretically suggested which were based on the stationary optical pumping of the atomic ground level during transits of atoms of a rarefied gas between walls of a thin gas cell, whose characteristic transverse dimension $D$ was much greater than its inner thickness $L$ [3-6]. Later, these methods were successfully realized experimentally for the precision spectral analysis of atoms [7-12] and for the laser frequency stabilization [12-15] on the basis of thin cells (with various inner thicknesses from 10 μm to 10 mm) containing $Cs$ [7-11, 13-15] or $Rb$ [12] vapors. It is important to note that, unlike the saturated absorption spectroscopy in usual "macroscopic" cells (where $L>>D$) [1], sub-Doppler spectroscopy in thin gas cells avoids crossover resonances and corresponding stabilization systems are much less affected by frequency fluctuations of the pumping radiation (which may be even broadband) because the velocity selection of optically pumped atoms originates from the cell geometry. In fact, a collection of optically pumped atoms (molecules) in the thin cell is the compact analog of the atomic (molecular) beam. The divergence of such a beam is determined, in particular, by the small ratio $L/D<<1$ of the cell inner thickness $L$ to its transversal dimension $D$. Papers [16,17] present reviews on methods, achievements and possibilities of this sub-Doppler spectroscopy in thin gas cells.

In particular, the effective spectroscopy method was elaborated with pumping and probe light beams, which may travel through a thin cell even in orthogonal directions [6, 9-14]. For example, authors of experimental works [9-14] used cylindrical cells (containing $Cs$ or $Rb$ vapor) with the diameter $D$ about 30 mm and different inner thickness $L<D$ from 0.5 mm to 10 mm. In this case the relaxation of pumped long-lived (ground or metastable) atomic levels is caused mainly by collisions of atoms with the end walls of the cell. Therefore the optical pumping and transit time effects in the thin gas cell can produce a nonequilibrium distribution of atoms on the velocity projection $\left|v_z\right|$ along the cell thickness (axis $z$) for these atomic levels in the region of sufficiently small values $\left|v_z\right|$. Thus sub-Doppler resonances appear in the absorption spectrum of the probe monochromatic wave (running along the $z$ axis) on central frequencies of optical transitions from given pumped long-lived states to any excited atomic quantum levels. These resonances may be observed directly by the record of the difference between probe signals with and without pumping radiation [9-14]. In particular the effective frequency stabilization of a diode laser was achieved on hyperfine components of the $Cs\,D_2$ line on the basis of given thin cell spectroscopic methods [13, 14]. Afterwards, the corresponding US Patent on the laser frequency stabilization device and the method for the laser frequency stabilization was received [18].

Authors of the brief communication [19] informed on their experimental demonstration of a possible rise of sub-Doppler spectroscopy resolution in thin gas cells by means of the spatial separation of pump and probe radiations. Detailed theoretical analysis of such a spectroscopy resolution rise was carried out in the following article [20] for the stationary optical pumping. Then authors of the paper [21] also confirmed the essential narrowing of the absorption sub-Doppler resonances by the spatial separation of unidirectional pump and probe laser beams in experiments with a 1mm-long $Rb$ vapor cell. Moreover these scientists suggested and realized the new method of sub-Doppler spectroscopy (with the same 1mm-long $Rb$ cell), based on the starting optical pumping of the ground $Rb$ term by the pulse of the monochromatic radiation [21]. In the definite time $t$ after the action of this pulse, the thin vapor cell was probed by the comparatively weak light pulse with the same frequency and direction (from the same diode laser). Thus authors of the paper [21] have revealed the contribution of comparatively slow-speed atoms with the pumped ground term, which had not time to undergo collisions with walls of the cell during the time $t$.

However, comparatively weak absorption signals in thin gas cells reduce their effectiveness. Therefore, for the most important case of spatially separated pumping and probe light beams, we suggest a new cell, which contains a series of successive plane-parallel thin gas layers (Fig.1). According to Fig.1, this macroscopic cell with a rarefied gas contains the selector 2 of velocities for optically pumped atoms. The sufficiently weak monochromatic probe beam travels (in the z direction) through the transparent opening 3. At the same time, the spatially separated pumping beam (which may be even broadband) irradiates the region of the gas cell 1 out of the selector. The selector material 2 should not be chemically active with atoms of the gas medium and may be even nontransparent for incident pumping and probe radiations. Under given conditions, atoms with the optically pumped ground quantum term may freely fly (without their collisional relaxation) from the pumping region 1 to the probe region 3 only through sufficiently thin splits, which are cut out in the selector material 2 in planes orthogonal with respect to the axis z. Thus it is possible to receive (in the probe region) a number of plane-parallel thin gas layers of optically pumped atoms (with very small velocity projections $\left|v_z\right|$) in limits of one macroscopic gas cell. Such a cell may be easier fabricated and operated at lower temperatures (for a sufficient gas concentration) in comparison with a single gas cell with a submillimeter inner thickness. The difference absorption signal of the monochromatic probe beam may be detected in experiments (that is with and without spatially separated pumping radiation). Then, in the suggested scheme, the narrowing of the Doppler broadening of a recorded spectral line will be determined by a large ratio of the characteristic extent $0.5\left(D_1 - D_2\right)$ of a split to its small thickness $\Delta l$ (Fig.1a). For example, it is possible to cut out sufficiently many such plane parallel splits with thickness $\Delta l \sim 0.1$mm in the selector material for a sufficiently compact gas cell with the diameter $D$ and length $L$ of the order of 1cm.

In this work we analyze the velocity distributions of optically pumped atoms in the probe region 3 under action of a broadband pumping radiation in the spatially separated region 1 of the suggested multilayer gas cell (Fig.1). Corresponding analytical calculations have been carried out for an arbitrary time dependence of the optical pumping rate (section 2). Following detailed investigations are carried out for the stationary pumping radiation and also after its abrupt interruption, when selection of more slow-speed pumped atoms may be achieved (section 3). Possibilities of sub-Doppler spectroscopy of atoms and molecules on the basis of given multilayer gas cells are discussed.

## 2. Basic relationships

Let us consider the optical pumping of a gas medium by the homogeneous broadband pumping radiation in the coaxial region of the cylindrical cell (Fig.1), corresponding to the radius interval $0.5D_1 \leq r \leq 0.5D$. This optical radiation excites atoms (or molecules) from the sublevel $a$ of the ground atomic term to the state $b$ on the resonant nonclosed quantum transition $a \Rightarrow b$. It is assumed that the intensity of the given radiation is not too large, when the population of the upper level $b$ is negligible in comparison with the population of the lower state $a$. The sufficiently rarefied gas medium in the cell is considered, when an interaction between atoms is negligible. We also don't take into account light pressure effects on atomic particles. Then known atomic density matrix equations [1] yield the following equation for the population $\rho_a(\mathbf{R},\mathbf{v},t)$ of atoms on the long-lived level $a$:

$$\frac{\partial \rho_a}{\partial t} + \mathbf{v}\frac{\partial \rho_a}{\partial \mathbf{R}} = -W(t)\eta(0.5D-r)\eta(r-0.5D_1)\rho_a , \qquad (1)$$

where $t$ is the time, $\mathbf{v}$ and $\mathbf{R}$ are the atomic velocity and coordinate vector respectively, $W(t)=\xi_{ab}(t)(1-B_{ba})$ is the optical pumping rate, $\xi_{ab}(t)$ is the probability of the atomic excitation by the pumping radiation on the nonclosed transition $a \Rightarrow b$, $B_{ba}$ is the probability of the subsequent radiative decay on the channel $b \rightarrow a$ ($B_{ba} <1$), $\eta(x)$ is the step function ($\eta(x)=1$ if $x \geq 0$ and $\eta(x)=0$ if $x < 0$). Eq.(1) must be supplemented by boundary conditions which depend on features of atomic collisions with walls of the cell and its velocity selector (Fig.1). We will assume that the equilibrium distribution for both atomic velocities and populations of quantum levels are established due to such collisions. The population $\Delta\rho_a$ of optically pumped atoms (from the long-lived level $a$) is determined by the formula:

$$\Delta\rho_a(\mathbf{R},\mathbf{v},t)=\sigma_a F(\mathbf{v})-\rho_a(\mathbf{R},\mathbf{v},t), \qquad (2)$$

where $\sigma_a$ is the equilibrium density of atoms on the level $a$ with the Maxwell distribution $F(\mathbf{v})$. We will consider the multilayer gas cell (Fig.1), where a large number $N>>1$ of similar equidistant splits (with the same thickness $\Delta l$) are cut of in the cylindrical velocity selector (Fig.1). Under given conditions, we receive from Eqs.(1), (2) the following expression for the population $\Delta\rho_a^{(n)}$ of optically pumped atoms (from the level $a$), which achieve the central axis $z$ of the cylindrical cell (Fig.1) without collisional relaxation through a thin split with a number $n$ ($1 \leq n \leq N$) and corresponding coordinates $z_n$ and $(z_n + \Delta l)$ along the axis $z$:

$$\Delta\rho_a^{(n)}(z, v_z, v_r, t) = \sigma_a F_l(v_z) F_r(v_r) \left\{ 1 - \exp\left[ -\int_{(t-0.5D/v_r)}^{(t-0.5D_1/v_r)} W(t')dt' \right] \right\} \left\{ \eta\left( z_n + \Delta l - z + \frac{D_2 v_z}{2v_r} \right) \right.$$

$$\times \eta\left( z - z_n - \frac{D_1 v_z}{2v_r} \right)\eta(v_z) + \eta\left( z_n + \Delta l - z + \frac{D_1 v_z}{2v_r} \right)\eta\left( z - z_n - \frac{D_2 v_z}{2v_r} \right)\eta(-v_z) \right\}. \quad (3)$$

In this Eq.(3) $v_z$ and $v_r$ are longitudinal and radial components of the atomic velocity $\mathbf{v}$, respectively, characterized by Maxwell distributions:

$$F_l(v_z) = \pi^{-1/2} u^{-1} \exp(-v_z^2 u^{-2}), \; F_r(v_r) = 2v_r u^{-2}\exp(-v_r^2 u^{-2}), \quad (4)$$

where $u$ is the most probable atomic speed in the gas. The average population $\Pi_a(v_z, v_r, t)$ of optically pumped atoms on the whole central axis $z$ of the cell (Fig.1) is determined from Eq.(3):

$$\Pi_a(v_z, v_r, t) = L^{-1} \sum_{n=1}^{N} \int_0^L \Delta\rho_a^{(n)}(z, v_z, v_r, t)dz, \quad (5)$$

We will assume that the relationship $\Delta l << [1 - (D_2/D_1)]L$ takes place for geometrical parameters of the cell and its velocity selector (Fig.1). Then we receive from Eqs.(3)-(5) the following function $\Pi_a(v_z, v_r, t)$, which approximately is the sum of equal contributions from all equidistant $N >> 1$ splits:

$$\Pi_a(v_z, v_r, t) = \sigma_a \left( \frac{N\Delta l}{L} \right) F_l(v_z) F_r(v_r) \left\{ 1 - \exp\left[ -\int_{(t-0.5D/v_r)}^{(t-0.5D_1/v_r)} W(t')dt' \right] \right\}\left[ 1 - \frac{(D_1 - D_2)}{2\Delta l}\frac{|v_z|}{v_r} \right]$$

$$\times \eta\left[ 1 - \frac{(D_1 - D_2)}{2\Delta l}\frac{|v_z|}{v_r} \right]. \quad (6)$$

Now we may determine the effective population $P_a(v_z, t)$ of optically pumped atoms on the central axis $z$ of the cell versus their longitudinal velocity projection $v_z$ at a moment $t$:

$$P_a(v_z, t) = \int_0^\infty \Pi_a(v_z, v_r, t)dv_r. \quad (7)$$

In case of the stationary optical pumping with the sufficiently high constant rate $W = W_0 >> u/(D - D_1)$, it is possible to receive from Eqs. (6), (7) the following formula for the limitary value $P_a(v_z)$:

$$P_a^{(s)}(v_z) \approx \sigma_a \left( \frac{N\Delta l}{L} \right) F_l(v_z) \left\{ \exp\left[ -\left( \frac{(D_1 - D_2)v_z}{2\Delta l u} \right)^2 \right] + \left[ \frac{(D_1 - D_2)|v_z|}{2\pi^{0.5}\Delta l u} \right]\left[ erf\left( \frac{(D_1 - D_2)|v_z|}{2\Delta l u} \right) - 1 \right] \right\}, \quad (8)$$

where $erf(x) = 2\pi^{-0.5}\int_0^x \exp(-y^2)dy$ is the error function. We note, that the "saturated" stationary velocity distribution $P_a^{(s)}(v_z)$ in Eq.(8) does not depend on the pumping rate $W = W_0$. Indeed, in the limit $W_0(D-D_1)/u \to \infty$ the repumping of all atoms occurs from the level $a$ (on the resonance nonclosed transition $a \Rightarrow b$) during transits of these atoms through the pumping region 1 of the cell (Fig.1). If the similar situation takes place for sufficiently intensive nonstationary (in particular pulsed) optical pumping to the moment $t$=0 of its interruption, then we receive from Eqs. (6),(7) the following time dependence for the velocity distribution $P_a(v_z, t)$ of optically pumped atoms when $t$>0:

$$P_a^{(s)}(v_z, t) \approx \sigma_a \left( \frac{N \Delta l}{L} \right) F_l(v_z) \left\{ \exp\left[ -\left( \frac{(D_1 - D_2)v_z}{2\Delta l u} \right)^2 \right] - \exp\left( -\frac{D^2}{4u^2 t^2} \right) + \left[ \frac{(D_1 - D_2)|v_z|}{2\pi^{0.5}\Delta l u} \right] \right.$$

$$\times \left. \left[ erf\left( \frac{(D_1 - D_2)|v_z|}{2\Delta l u} \right) - erf\left( \frac{D}{2ut} \right) \right] \right\} \eta \left[ \frac{D}{2t} - \frac{(D_1 - D_2)}{2\Delta l} |v_z| \right]. \tag{9}$$

Further we will present values $P_a(v_z, t)$ calculated on the basis of Eq.(7) in units of the maximum value $A_0 = P_a^{(s)}(v_z = 0)$ from Eq.(8) that is

$$A_0 = \frac{\sigma_a}{\pi^{0.5} u} \left( \frac{N \Delta l}{L} \right). \tag{10}$$

## 3. Discussion of calculation results

Let us, at first, consider the case of the stationary optical pumping with the constant rate $W(t) = W_0$. Then, according to Eq.(6), at a fixed longitudinal velocity projection $v_z$, the effective distribution $\Pi_a(v_z, v_r)$ of optically pumped atoms on their radial velocity component $v_r$ is determined by the most probable speed $u$ of atoms in the gas cell. At the same time, the effective stationary distribution $P_a(v_z)$ [Eq.(7)] of given atoms on the velocity projection $v_z$ is the narrow peak centered on the value $v_z$=0 (curves 1-3 in Fig.2). The half-width $\Delta v_z$ on the half-height of this function $P_a(v_z)$ is much less than the most probable atomic speed $u$ and decreases at rise of the geometrical ratio $0.5(D_1 - D_2)/\Delta l$ of a split in the velocity selector. Indeed, according to Eq.(6), only definite optically pumped atoms may arrive to the central axis $z$ of the cell (Fig.1) without collisional relaxation from its pumping region 1, whose velocity components $v_z$ and $v_r$ satisfy the condition $|v_z| \leq [2\Delta l /(D_1 - D_2)] v_r$.

Fig.3 shows the amplitude $A = P_a(v_z = 0)$ and the characteristic half-width $\Delta v_z$ of the stationary velocity distribution $P_a(v_z)$ [Eq.(7)] versus the optical pumping rate $W_0$ for different ratios $0.5(D_1 - D_2)/\Delta l$. Growth of the pumping intensity leads to rise both the width and amplitude of this distribution. We note that the amplitude $A$ does not directly depend on the geometrical factor $0.5(D_1 - D_2)/\Delta l$ and asymptotically approaches to the maximum value $A_0$[Eq.(10)] (Fig.3a). At the same time, according to the "saturated" distribution $P_a^{(s)}(v_z)$ from Eq.(8), the half-width $\Delta v_z$ at rise of the pumping rate $W = W_0$ tends to the value $\Delta v_z^* \approx [0.7\Delta l /(D_1 - D_2)]u << u$. At fall of the stationary pumping intensity, the half-width $\Delta v_z$ may be decreased from this maximum $\Delta v_z^*$ up to the value close to $0.5\,\Delta v_z^*$ (Fig.3b). However, for such low pumping rate $W_0 << u/(D - D_1)$ the corresponding amplitude $A$ of the stationary velocity distribution $P_a(v_z)$ is too small (Fig.3a).

Let us consider also dynamics of velocity distributions of optically pumped atoms in the central (probe) region of the cell (Fig.1) after the abrupt interruption of the stationary pumping at the moment $t=0$. In this case the time dependence of the pumping rate $W(t) = W_0\eta(-t)$ is the step function. Curves 3-5 in Fig.2 shows corresponding decrease of the number of optically pumped atoms in course of the time $t>0$. In consequence the selection of more slow speed pump atoms may be achieved. Indeed, according to Eq.(6), at the pumping rate $W(t) = W_0\eta(-t)$ following restrictions take place for velocity components $v_r$ and $v_z$ of given atoms on the central axis $z$ of the cell at a moment $t>0$:

$$v_r < \frac{D}{2t}, \qquad |v_z| < \frac{2\Delta l}{(D_1 - D_2)} v_r. \qquad (11)$$

We see from this Eq.(11), that decrease of radial components $v_r$ leads to decrease of corresponding longitudinal velocity projections $v_z$ for optically pumped atoms. In particular $v_r << u$ and $|v_z| \leq [\Delta l/(D_1 - D_2)](D/t)$ if $t >> 0.5(D/u)$. Fig.4 shows dynamics of the characteristic half-width $\Delta v_z(t)$ and the amplitude $A(t) = P_a(v_z = 0, t)$ of the velocity distribution $P_a(v_z, t)$ [Eq.(7)] for optically pumped atoms during the time $t>0$. Growth of the pumping intensity leads to increase of these values $\Delta v_z(t)$ and $A(t)$. Meanwhile, in case of the limitary intensive pumping we receive from Eq.(9) for time $t > (D/u)$ the half-width $\Delta v_z^*(t) \approx 0.29[\Delta l/(D_1 - D_2)](D/t)$, which does not depend on the pumping rate. The half-width $\Delta v_z(t)$ may asymptotically decrease up to zero value in the course of time $t>0$. However the more fast decrease of the amplitude $A \sim t^{-2}$ takes place in comparison with the corresponding half-width $\Delta v_z(t) \sim t^{-1}$ when $t > (D/u)$ (Fig.4).

## 4. Conclusions

We have established and analyzed narrow velocity distributions of atoms optically pumped on a nonclosed transition from a sublevel of the ground atomic term in the central (probe) region of the suggested multilayer gas cell (Fig.1). In particular, sufficiently small half-widths $\Delta v_z << u$ of such distributions on the longitudinal velocity projection $v_z$ of pumped atoms are shown in Figs.3 and 4. Sub-Doppler spectroscopy of given atoms may be realized by means of the probe monochromatic light beam running along the axis $z$ of such a cell (Fig.1). The difference absorption signal of this probe beam may be detected, that is with and without pumping radiation in the spatially separated region of the cell. Then, at the frequency scan of the probe light around the center $\omega_0$ of an investigated atomic transition $a \Rightarrow f$ from the pumped lower state $a$ to any excited quantum level $f$, the Doppler broadening of such an absorption signal will be determined by the value $(\omega_0/c)\Delta v_z$.

At the stationary optical pumping, the effective narrowing of the Doppler width is determined by the factor $(\Delta v_z/u) \leq 2\Delta l/(D_1 - D_2) << 1$, that is by the small ratio of characteristic sizes of thin splits in the velocity selector of the cell (Fig.1). The interruption of the pumping radiation at a moment $t = t_0$ allows to achieve more narrow velocity distributions of optically pumped atoms in the course of time $t > t_0$ (Figs.2, 4). However, for corresponding experiments and applications, it will be necessary to determine such a time interval $|t_2 - t_1|$ between moments $t_2 > t_0$ and $t_1 > t_0$, when detection of a sufficient absorption signal of the probe beam (by the difference method elaborated in the work [21]) still will be possible, in spite of decreasing number of pumped atoms during the time $t > t_0$.

Detected sub-Doppler absorption signal will be proportional to a number $N$ of thin splits in the velocity selector of the suggested cell (Fig.1). Therefore in case of sufficiently large their number $N \gg 1$, given cells may be used for the sub-Doppler spectroscopy not only atoms but also molecules with comparatively weak spectral lines.

We have considered the coaxial optical pumping by the broadband radiation for the cylindrical geometry of the multilayer gas cell (Fig.1). Meanwhile, obtained qualitative results are valid also for different shapes of a gas cell with spatially separated pumping and probe gas regions, which are connected by plane-parallel thin splits for velocity selection of optically pumped atoms. Moreover, the pumping may be realized also by a resonant narrow-band laser radiation directed even in a transverse direction with respect to the probe light beam as, for example, in experimental works [9-14] with a single thin gas cell.

By present time, successful experiments have been carried out on the ultra-high resolution spectroscopy of *Cs* atoms in the collimated atomic beam by means of the femtosecond laser optical frequency comb [22]. The similar use of the probe radiation from such a comb, operating in a broad spectral interval, will essentially extend a number of analyzed atomic (molecular) quantum transitions by the considered sub-Doppler spectroscopy method in multilayer gas cells.

The suggested gas cell (Fig.1) is the compact analog of many plane-parallel atomic (molecular) beams and may be used not only for ultra-high resolution spectroscopy of atoms and molecules but also as the basis for new compact optical frequency standards of high accuracy.

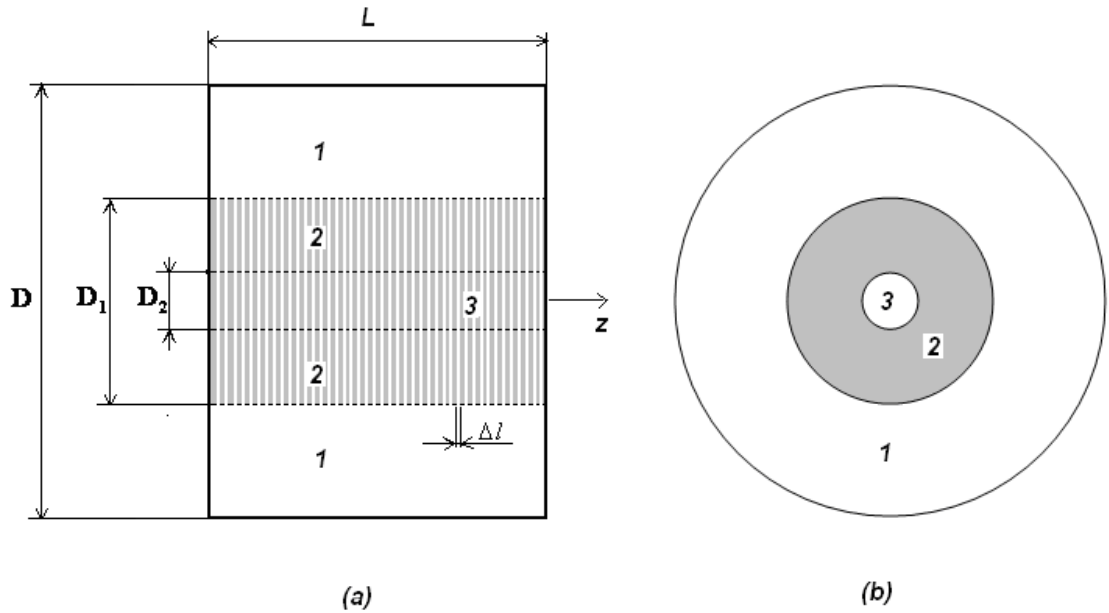

**Fig.1.** Scheme of the multilayer gas cell (with the diameter *D* and length *L*) in two orthogonal projections (a) and (b), which contains the velocity selector 2 for optically pumped atoms with the transparent opening 3 and a number of plane parallel splits (with the thickness $\Delta l$ ) between pumping region 1 and spatially separated probe region 3.

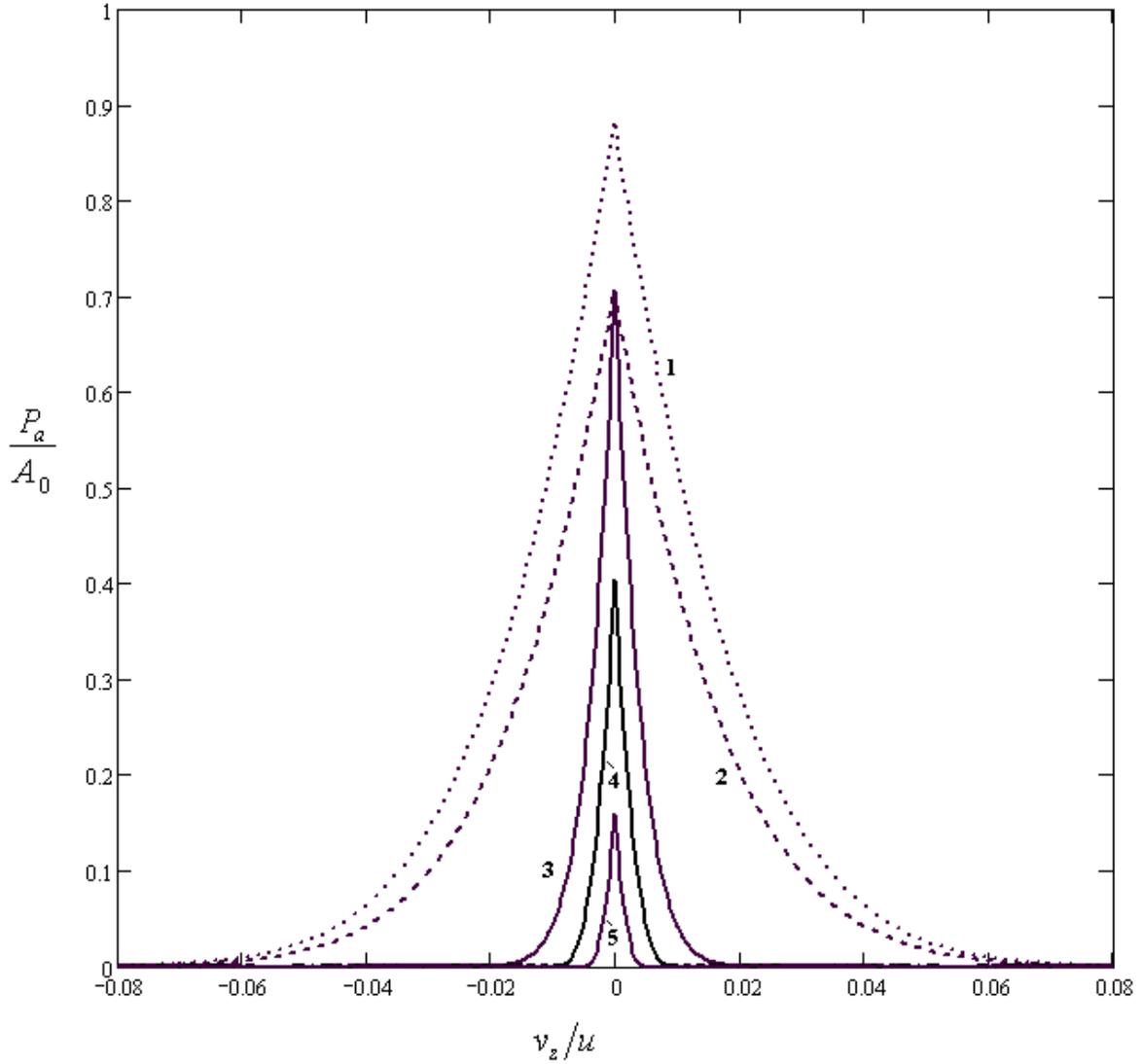

**Fig.2.** Distribution $P_a(v_z, t)$ of optically pumped atoms versus the longitudinal velocity projection $v_z$ on the central axis $z$ of the multilayer cell (Fig.1) in cases of the constant pumping rate $W(t) = W_0$ (curves 1-3) and after its step change $W(t) = W_0 \eta(-t)$ (curves 4, 5), when $0.5(D_1 - D_2)/\Delta l = (1,2)$ 25 and (3-5) 100, $0.5W_0(D - D_1)/u = (2$-$5)$ 1 and (1) 2, $t(D/2u) = (1$-$3)$ 0, (4) 1 and (5) 2.

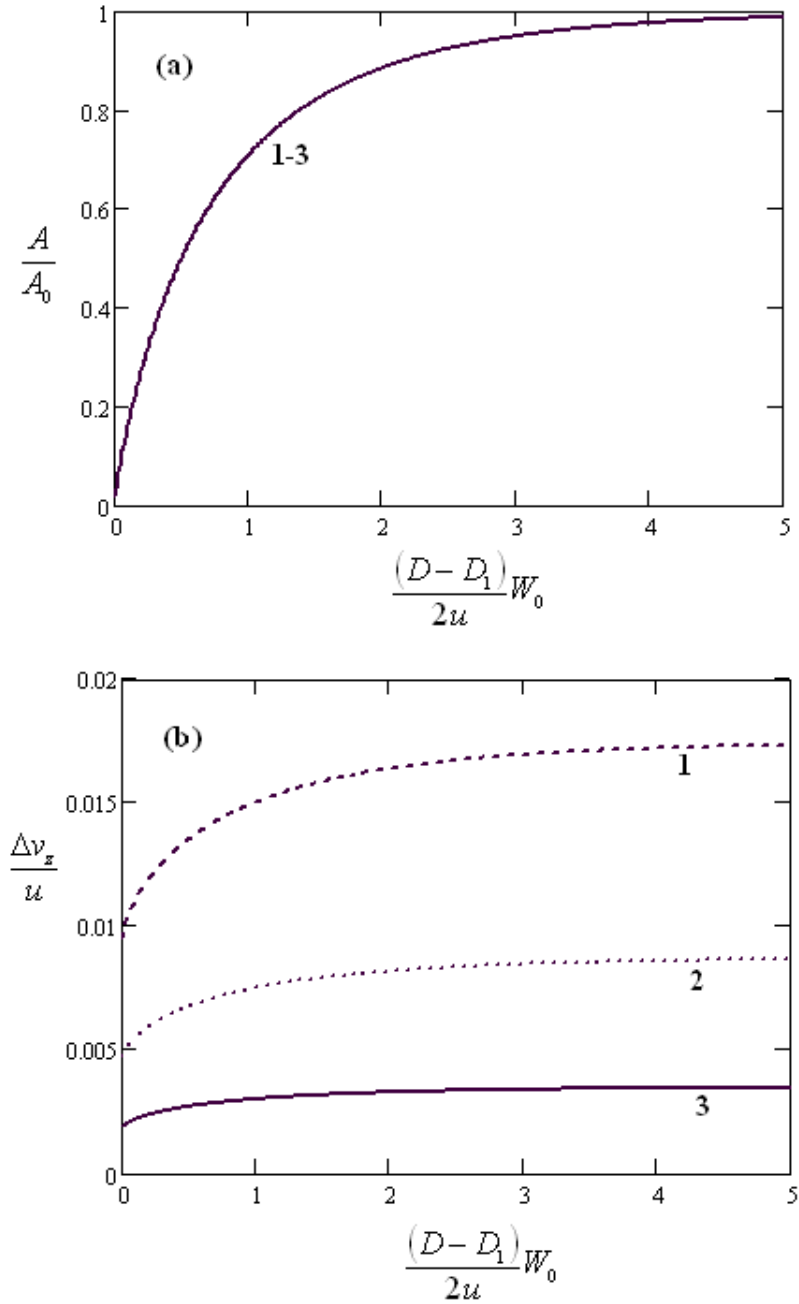

**Fig.3.** Dependence of the amplitude $A$ (a) and the half-width $\Delta v_z$ (b) of the stationary velocity distribution $P_a(v_z)$ of optically pumped atoms on the optical pumping rate $W_0$, when $0.5(D_1 - D_2)/\Delta I$ =(1) 20, (2) 40 and (3) 100.

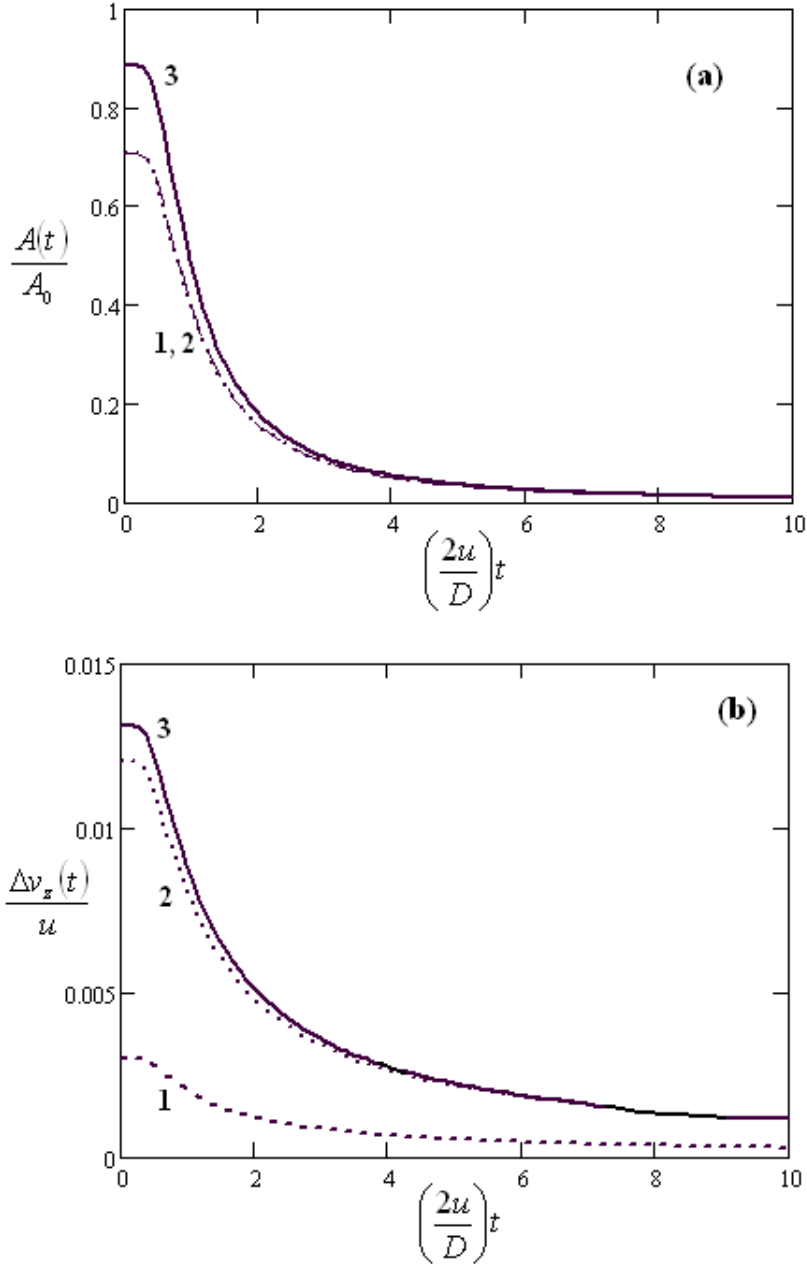

**Fig.4.** Time dependence of the amplitude $A(t)$ (a) and the half-width $\Delta v_z(t)$ (b) of the velocity distribution $P_a(v_z, t)$ of optically pumped atoms at the step change of the pumping rate $W(t) = W_0\eta(-t)$ if $0.5(D_1 - D_2)/\Delta l =$ (1) 100 and (2,3) 25, $0.5W_0(D - D_1)/u =$ (1,2) 1 and (3) 2.